\documentclass[aps, reprint, amsmath, amssymb, superscriptaddress]{revtex4-1} 

\usepackage{amsmath}
\usepackage{amsfonts}
\usepackage{amssymb}
\usepackage{amsthm}
\usepackage{fancyhdr}
\usepackage{enumerate}
\usepackage[top=1in, bottom=1in, left=1in, right=1in]{geometry}
\usepackage{graphicx}
\usepackage[labelfont=bf]{caption}

\usepackage{caption}
\usepackage{subcaption}

\usepackage{wasysym}
\usepackage{hyperref}

\usepackage{wrapfig}

\usepackage[toc]{appendix}

\begin{document}

\title{Takeover times for a simple model of network infection}

\author{Bertrand Ottino-L\"{o}ffler}
\affiliation{Center for Applied Mathematics, Cornell University, Ithaca, New York 14853}
\author{Jacob G. Scott }
\affiliation{Cleveland Clinic, Departments of Translational Hematology and Oncology Research and Radiation Oncology}
\author{ Steven H. Strogatz}
\affiliation{Center for Applied Mathematics, Cornell University, Ithaca, New York 14853}
\date{\today}

\begin{abstract}
We study a stochastic model of infection spreading on a network. At each time step a node is chosen at random, along with one of its neighbors. If the node is infected and the neighbor is susceptible, the neighbor becomes infected. How many time steps $T$ does it take to completely infect a network of $N$ nodes, starting from a single infected node? An analogy to the classic ``coupon collector" problem of probability theory  reveals that the takeover time $T$ is dominated by extremal behavior, either when there are only a few infected nodes near the start of the process or a few susceptible nodes near the end. We show that for $N \gg 1$, the takeover time $T$ is distributed as a Gumbel for the star graph; as the sum of two Gumbels for a complete graph and an Erd\H{o}s-R\'{e}nyi random graph; as a normal for a one-dimensional ring and a two-dimensional lattice; and as a family of intermediate skewed distributions for $d$-dimensional lattices with $d \ge 3$ (these distributions approach the sum of two Gumbels as $d$ approaches infinity). Connections to evolutionary dynamics, cancer, incubation periods of infectious diseases, first-passage percolation, and other spreading phenomena in biology and physics are discussed.
\end{abstract}

\maketitle

\section{Introduction}\label{Section_Intro}
Contagion is a topic of broad interdisciplinary interest. Originally studied in the context of infectious diseases \cite{anderson1991infectious, keeling2005networks, diekmann2012mathematical, pastor2015epidemic}, contagion has now been used as a metaphor for diverse processes that spread by contact between neighbors. Examples include the spread of fads and fashions \cite{bikhchandani1992theory, watts2002simple}, scientific ideas \cite{bettencourt2006power}, bank failures \cite{aharony1983contagion,allen2000financial, may2008complex,may2010systemic,haldane2011systemic}, computer viruses \cite{kephart1991directed}, gossip \cite{haas2006gossip}, rumors \cite{daley1965, draief2010epidemics}, and yawning \cite{provine2005yawning}. Closely related phenomena arise in probability theory and statistical physics in the setting of first-passage percolation \cite{aldous13, auffinger15}, and in evolutionary dynamics in connection with the spread of mutations through a resident population~\cite{lieberman05,antal06,hindersin14,hindersin15, ashcroft15}. We will use the language of contagion throughout, but bear in mind that everything could be reformulated in the language of the other fields mentioned above. 

In the simplest mathematical model of contagion, the members of the population can be in one of two states: susceptible or permanently infected. When a susceptible individual meets an infected one, the susceptible immediately becomes infected. Even in this idealized setting, interesting theoretical questions remain, whose answers could have significant real-world implications, as we will argue below. 

For example, consider the following model, motivated by cancer biology. Imagine a two-dimensional lattice of cells in a tissue, where each cell is either normal or mutated. At each time step a random cell is chosen, along with one of its neighbors, also chosen uniformly at random. If the first cell is mutated and its neighbor is normal, the mutated cell (which is assumed to reproduce much faster than its normal neighbor) makes a copy of itself that replaces the normal cell. In effect, the mutation has spread; it behaves as if it were an infection. This deliberately simplified model was introduced in 1972 to shed light on the growth and geometry of cancerous tumors~\cite{williams72}.

Here, we study this model on a variety of networks. Our question is: given a single infected node in a network of size $N$, how long does it take for the entire network to become infected? We call this the \emph{takeover time} $T$. It is conceptually related to the fixation time in population genetics, defined as the time for a fitter mutant to sweep through a  resident population. It is also reminiscent of the incubation period of an infectious disease, defined as the time lag between exposure to the pathogen and the appearance of symptoms; this lag presumably reflects the time needed for infection to sweep through a large fraction of the resident healthy cells. 

\begin{figure*}
\includegraphics[width = 0.95\textwidth]{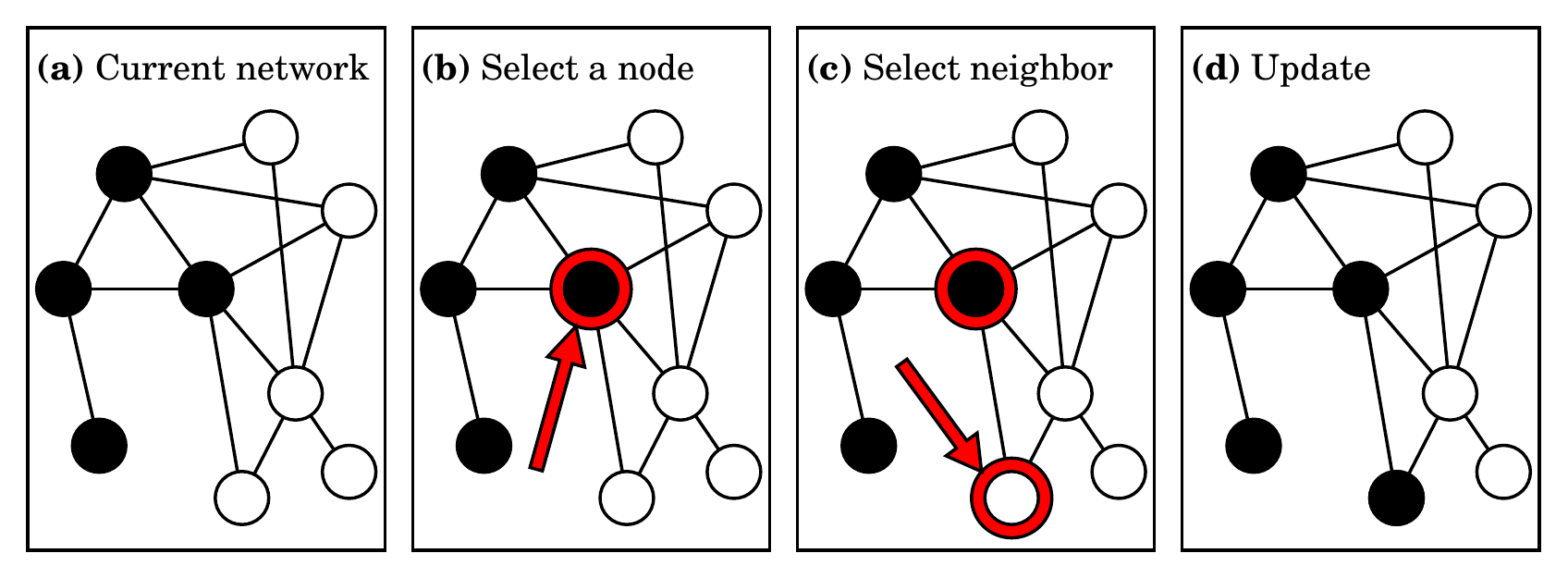} 
\caption{Simple model of infection spreading on a network. (a) A typical current state of the network is shown. Filled dots represent infected nodes, and open dots represent susceptible nodes. Links represent potential interactions. (b) At each time step, a random node is selected. (c) One of the node's neighbors is also selected at random. The infection spreads only if the node chosen in (b) happens to be infected and its neighbor chosen in (c) happens to be susceptible, as is the case here; then the state of the network is updated accordingly as in (d). Otherwise, if the node is not infected or the neighbor is not susceptible, nothing happens and the state of the network remains unchanged. In that case the time step is wasted.}
\label{SchematicPlot}
\end{figure*}

For the model studied here, the calculation of the network takeover time is inherently statistical because the dynamics are random. At each time step, we choose a random node in the network, along with one of its neighbors, also at random. If neither of the nodes are infected, nothing  happens and the time step is wasted. Likewise, if both are infected, the state of the network again doesn't change and the time step is wasted. Only if the first node is infected and its neighbor is susceptible does the infection progress, as shown in Figure~\ref{SchematicPlot}. 

The time course of the infection is interesting to contemplate. Intuitively, when the network is large, it seems that the dynamics should be very stochastic at first and take a long time to get rolling, because it is exceedingly unlikely that we will randomly pick the one infected node, given that there are so many other nodes to choose from. Similarly we expect a dramatic slowing down and enhancement of fluctuations in the endgame. When a big network is almost fully infected, it becomes increasingly difficult to find the last few susceptible individuals to infect.

These intuitions led us to suspect that the problem of calculating the distribution of takeover times might be amenable to the techniques used to study the classic ``coupon collector" problem in probability theory \cite{feller1968introduction, posfai10}. If you want to collect $N$ distinct coupons, and at each time step you are given one coupon at random (with replacement), what is the distribution of the time required to collect all the coupons? Like the endgame of the infection process,  the coupon collection process slows down and suffers large fluctuations when almost all the coupons are in hand and one is waiting in exasperation for that last coupon. Erd\H{o}s and R\'{e}nyi proved that for large $N$, the distribution of waiting times for the coupon collection problem approaches a Gumbel distribution~\cite{erdos61}. This type of distribution is right skewed, and is one of the three universal extreme value distributions \cite{fisher1928limiting, kotz2000extreme}. 

In what follows, we will show that for $N \gg 1$, the takeover time $T$ is distributed as a Gumbel for the star graph, and as the sum of two Gumbels for a complete graph and an Erd\H{o}s-R\'{e}nyi random graph. For $d$-dimensional cubic lattices, the dependence on $d$ is intriguing:  we find that $T$ is normally distributed for $d =1$ and $d=2$, then becomes skewed for $d \ge 3$ and approaches the sum of two Gumbels as $d$ approaches infinity. We conclude by discussing the many simplifications in our model, with the aim of showing how the model relates to more realistic models. We also discuss the possible relevance of our results to fixation times in evolutionary dynamics, population genetics, and cancer biology, and to the longstanding (yet theoretically unexplained) clinical observation that incubation periods for infectious diseases frequently have right-skewed distributions.

\section{One-dimensional Lattice} \label{Section_1D}

We start with a one-dimensional (1D) lattice. In this paper, we will always take lattices to have periodic boundary conditions, so imagine $N$ nodes arranged into a ring.

Suppose that $m$ nodes are currently infected. Let $p_m$ denote the probability that a susceptible node gets infected in the next time step. Notice that for a more complicated graph, $p_m$ might not be a well-defined concept, because it could depend on more than $m$ alone: the probability of infecting a new node could depend on the positions of the currently infected nodes, as well as on the susceptible node being considered. In such cases, we would need to know the entire current state of the network, not just the value of $m$, to calculate  the probability that the infection will spread. 

The 1D lattice, however, is especially tractable. Assuming that only one node is infected initially, at later times the infected nodes are guaranteed to form a contiguous chain. So for this simple case the graph state is indeed determined by $m$ alone. The only places where the infection can spread are from the two ends of the infected chain. (Even on more complicated networks, the dynamics of our model imply that the infected nodes always form contiguous regions, but few are as simple as this.) 

The spread of infection involves two events. First, the node chosen at random must lie on on the boundary of the infected cluster. Then, one of its neighbors that happens to be susceptible must be picked. So 
\begin{alignat}{1}
p_m =& \left(\mbox{Prob. of selecting node on boundary} \right) \notag \\
& \times \left(\mbox{Prob. of selecting susceptible neighbor} \right). \label{pmLogic}
\end{alignat}
Hence, for the ring, the probability that the infection spreads on the next time step reduces to $p_m = (2/N)\times(1/2) = 1/N$ for all $m$. 

\begin{figure}
\includegraphics[width = 0.5\textwidth]{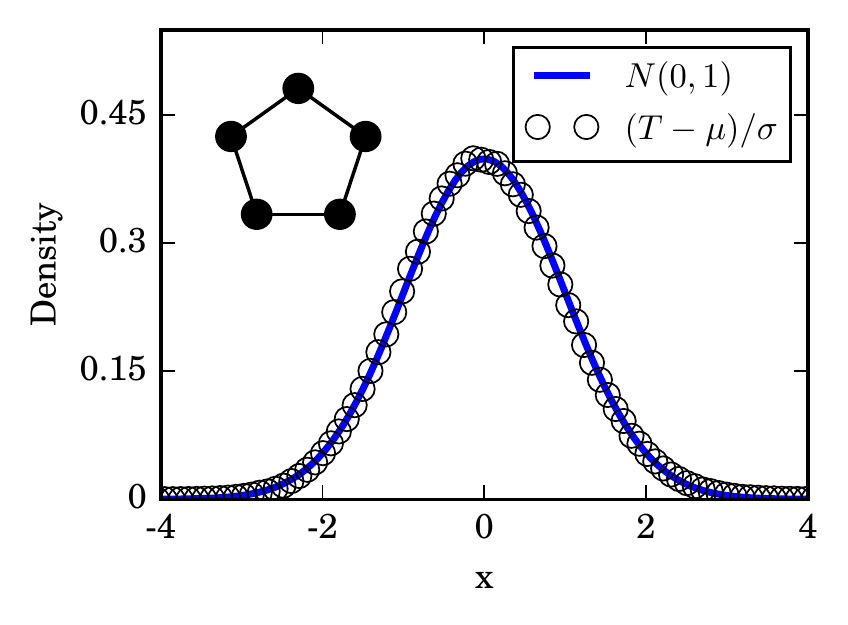}
\caption{Distribution of takeover times for 1D lattices with $N = 750$ nodes, obtained from $10^6$ simulations. The mean takeover time is $\mu= N(N-1)$ and its variance is $\sigma^2 = N(N-1)^2$, both found analytically. The simulation results are well approximated by a normal distribution, as expected. The diagram in upper left schematically shows a 1D lattice.}
\label{1DLatticePlot}
\end{figure}

Next, define the random variable $X_m = X(p_m)$ as the number of time steps during which the network has  exactly $m$ infected nodes. The probability that this state lasts for $k$ time steps is then given by
\begin{equation*}
P(X_m = k) = q_{m}^{k-1} p_m, 
\end{equation*}
for $k = 1, 2,\ldots,$ where $q_m := 1 - p_m$. To see this, note that $P(X_m = k)$ is the probability that no new infection occurs on the first $k-1$ steps, times the probability that infection does occur on step $k$.

Thus, for any network where $p_m$ is well defined, the time spent with $m$ infected nodes is a geometric random variable, with mean $1/p_m$ and variance $1/p_{m}^2 - 1/p_m$. In particular, since the ring has $p_m = 1/N$ for all $m$, we find that $X_m$ has mean $N$ and variance $N^2-N$ in this  case.

The takeover time for any network is $$T = \sum_{m=1}^{N-1} X_m,$$ the sum of all the individual times required to go from $m$ to $m+1$ infected nodes, for $m=1, \ldots, N-1$. (Equality, in this case, means equality in distribution, as it will for all the other random variables considered throughout this paper.) 

In the case of the 1D lattice, all the $X_m$ are identical. However, their means and variances depend on $N$, which prevents us from invoking the usual Central Limit Theorem to deduce the limiting distribution of $T$. However, we can invoke a generalization of it known as the Lindeberg-Feller theorem. See Appendix~\ref{RingLindeberg} for more details. 

After normalizing $T$ by its mean, $\mu= N(N-1)$, and its standard deviation, $\sigma = (N-1)\sqrt{N}$, we find
\begin{equation} \label{1DLatticeLimit}
\frac{T - N(N-1)}{(N-1)\sqrt{N}} \xrightarrow{d} \mbox{Normal}(0,1),
\end{equation}
where the symbol $\xrightarrow{d}$ means convergence in distribution as $N$ gets large. Figure~\ref{1DLatticePlot} confirms that the takeover times are normally distributed in the limit of large rings.

\section{Star graph} \label{Section_Star}

\begin{figure}
\includegraphics[width = 0.5\textwidth]{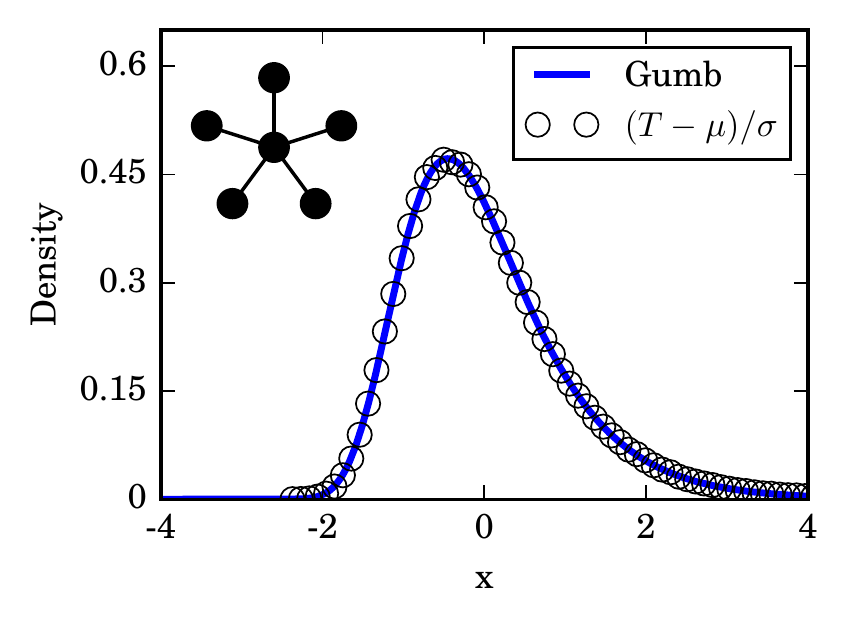}
\caption{Distribution of takeover times for a star graph with $N = 120$ spoke nodes, obtained from $10^6$ simulation runs. The mean $\mu$ and characteristic width $L$ are given by $\mu= (N+1)(N)\sum_{m=1}^{N}1/m$ and $L = N(N+1)$. The numerically generated histogram of takeover times closely follows the predicted Gumbel distribution, even for the small $N$ used here. The schematic diagram in upper left shows a star network.}
\label{StarPlot}
\end{figure}

A star graph is another common example for infection models. Here, $N$ separate ``spoke'' nodes all connect to a single ``hub'' node and to no others, as illustrated in the upper left of Figure~\ref{StarPlot}. We will assume the initial infection starts at the hub, since starting it in a spoke node would require only a trivial adjustment to the calculations below.

Let $m$ be the number of spoke nodes that are currently infected. As in the ring case, $p_m$ (the probability to go from $m$ to $m+1$ in the next time step) is a well-defined quantity that depends on $m$ alone, and not on any other details of the network state. Using the logic of Eq.~\eqref{pmLogic}, we get 
\begin{equation} \label{star_pm}
p_m = \frac{1}{N+1}\cdot \frac{N-m}{N}
\end{equation}
for $m = 0, 1, \ldots, N-1$. Here, $1/(N+1)$ is the probability of choosing the infected hub as the first node, and $(N-m)/N$ is the probability of selecting one of the $N-m$ currently susceptible spoke nodes, out of the $N$ spoke nodes in total, as its neighbor. 

Now that $p_m$ is in hand for the star graph, we can define the random variable $X_m$ and the takeover time $T$ just as we did for the one-dimensional ring. The only difference is that the $m$-dependence of $p_m$ is now controlled entirely by the factor $(N-m)/N$. 

That same factor turns up in a classic probability puzzle called the {\it coupon collector's problem}~\cite{feller1968introduction, posfai10}. At the time of this writing, millions of children are experiencing it firsthand as they desperately try to complete their collection of pocket-monsters in the Pok\'{e}mon videogame series.  

To see the connection, suppose you are trying to collect $N$ distinct items, and you have $m$ of them so far. If you are given one of the $N$ items at random (with replacement), the probability it is new to your collection is $(N-m)/N$, the same factor we saw above, and  precisely analogous to the probability $p_m$ of adding a new node to the infected set. Likewise, the waiting time to collect all $N$ items is precisely analogous to the time $T$ needed to take over the whole star graph.  The only difference is the constant factor $1/(N+1)$ in Eq.~\eqref{star_pm}.

The limiting distribution of the waiting time for the coupon collector's problem is well known. Although it  resembles a lognormal distribution~\cite{read98}, in fact it is a Gumbel distribution in the limit of large $N$, given the right scaling ~\cite{erdos61, posfai10, rubin65, baum65}. We will now show that the same is true for our problem.

The first move is to approximate the geometric random variables $X(p_m)$ by exponential random variables $\mathcal{E}(p_m)$, with density
\begin{equation*}
P(\mathcal{E}(p_m) = x)dx = p_m e^{-p_m x} dx; x \geq 0. 
\end{equation*}
From here we define the random variable $F := \sum_{m=0}^{N-1} \mathcal{E}(p_m)$, which has mean $\mu = \sum_{m=0}^{N-1} 1/p_m$. 

It can be shown (see Appendix~\ref{ConvergeToExp}) that for a large class of $p_m$ and normalizing factors $L := L(N)$ that
\begin{equation} 
\frac{T - \mu}{L} \sim \frac{F - \mu}{L},
\end{equation}
where the symbol ``$\sim$'' means the ratio of characteristic functions goes to 1 as $N$ gets large. That is, the random variables on both sides converge to each other in distribution as $N$ gets large. 

In the traditional coupon collector's problem we would take $L = N$; but because of that $(N+1)$ factor, what we want is $L = N(N+1)$. Thanks to the fact we are now using exponential variables, we now know
\[F/L = \sum_{m=0}^{N-1} \mathcal{E}(p_m/L) = \sum_{k=1}^{N} \mathcal{E}(k) \]
(using $k = N - m = 1, 2, \ldots, N$). A nice closed form for the distribution of the sum of a collection of distinct exponential variables is known~\cite{ashcroft15, baum65}, but for convenience's sake we rederive it in Appendix~\ref{ExpSum}. When we have any finite $N$ we can simply write the relevant distribution function $g_N(x)$ for $x \geq 0$ as
\begin{equation*}
g_{N}(x) = \sum_{k=1}^N k e^{-kx} \prod_{r\not= k}^N \frac{r}{r- k},
\end{equation*} 
which can be manipulated into 
\begin{equation*}
g_{N}(x) = N e^{-x} (1 - e^{-x})^{N-1}.
\end{equation*}
From here, we can find the distribution for $(F - \mu)/L$, and $(T - \mu)/L$ by extension. Therefore, taking the limit of large $N$ and using the standard approximation of the harmonic sum for $\mu$ gives us
\begin{equation} \label{StarDist}
f(x) = e^{-(x+\gamma)} \exp\left(- e^{-(x+\gamma)} \right)
\end{equation}
as the density, where $\gamma \approx 0.5772$ is the Euler-Mascheroni constant. The density in \eqref{StarDist} is a special case of the Gumbel distribution, denoted $\mathrm{Gumbel}(\alpha, \beta)$ and defined to have the density 
\begin{equation}\label{GenGumbel}
h(x) = \beta^{-1}  e^{-(x-\alpha)/\beta}\exp\left(- e^{-(x-\alpha)/\beta} \right).
\end{equation}
Specifically, we find
\begin{equation}
\frac{T - \mu}{N(N+1)} \xrightarrow{d} G,
\end{equation}
where $G$ is a Gumbel random variable distributed according to $\mathrm{Gumbel}(-\gamma, 1)$. 

This distribution can be tested against simulation, and it works nicely as seen in Figure~\ref{StarPlot}. Gumbel distributions have arisen previously in infection and birth-death models \cite{ashcroft15, gautreau07, williams65}, and are well known in extreme-value theory~\cite{feller1968introduction, fisher1928limiting, kotz2000extreme}, but the fact that they show up here as a result of a network topology is unexpected.

\section{Complete graph} \label{Section_Complete}
The complete graph on $N$ nodes corresponds to a ``well-mixed population'' and is one of the most common topologies in infection models. This network consists of $N$ mutually connected nodes, so the location of the initial infection does not matter. 

Given $m$ infected nodes, we once again have a well-defined $p_m$. Using the concept behind Eq.~\eqref{pmLogic}, we find 
\begin{equation} \label{pm_complete}
p_m = \frac{m}{N}\left(1- \frac{m-1}{N-1}\right) = \frac{m}{N}\cdot \frac{N-m}{N-1} 
\end{equation} 
for $m = 1, \ldots, N-1$. For the sake of convenience, we will collect these probabilities into a vector $p = (p_m)_{m=1}^N$. As in the case of the star graph, we can approximate the takeover time $T$ by summing exponential random variables instead of geometric ones. So

\begin{alignat}{1}\label{Comp-GeoToExp}
\frac{T - \mu}{N} \sim \sum_{m=1}^{N-1} \frac{\mathcal{E}(p_m) - 1/p_m}{N} =: S(p).
\end{alignat}
To compress notation, we defined $S(p)$ to be the normalized sum of exponential random variables across the entries of the vector $p$. 

The specific $p$ in Eq.~\eqref{pm_complete} has some helpful symmetry. Notice that if $k = N - m$, then 
\begin{equation*}
p_k = \frac{k}{N}\cdot \frac{N-k}{N-1} = \frac{N-m}{N}\cdot \frac{m}{N-1} = p_m. 
\end{equation*}
This symmetry means that the second half of the takeover looks just like the first half played backwards. If we set $p^{(f)}$ to be the front half of the $p$-vector and $p^{(b)}$ to be the back half of $p$, then we know 
\begin{equation}\label{Comp-GoHalfsies}
S(p) = S\left(p^{(f)}\right) + S\left(p^{(b)}\right) .
\end{equation}

Because we have a symmetry and the order we add the individual exponential variables will not matter, the random variables $S(p^{(f)})$ and $S(p^{(b)})$ should be equal in distribution. Although $N$ being odd or even may seem to be distinct cases, we will find that the distinction does not matter.

\begin{figure}
\includegraphics[width = 0.5\textwidth]{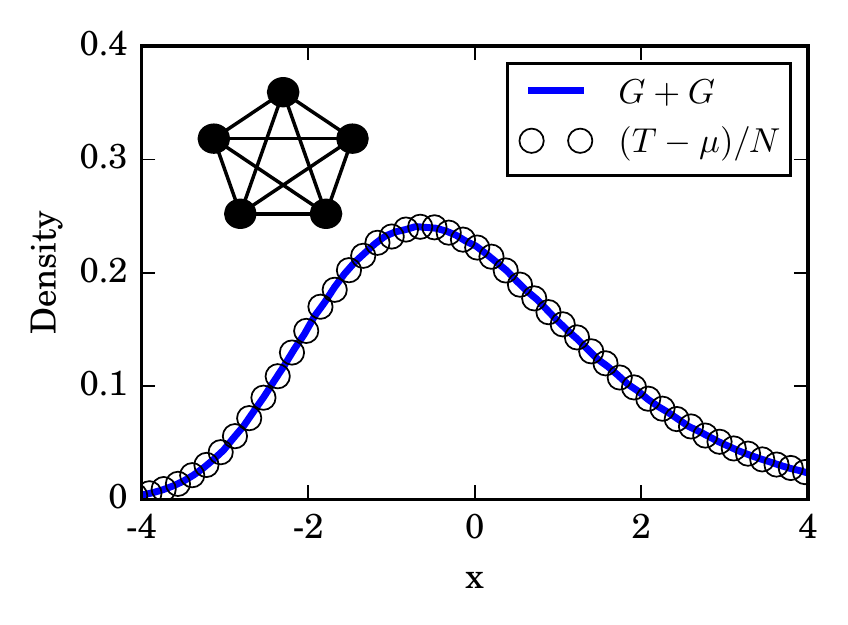}
\caption{Distribution of takeover times for a complete graph with $N = 450$ nodes. The histogram is based on $10^6$ simulation runs. The mean takeover time is $\mu = \sum_{m=1}^{N-1} 1/p_m$ exactly. Here, the numerically generated distribution fits closely to the convolution of two Gumbel distributions, produced using $5 \times 10^6$ samples. The schematic diagram in the upper left shows a complete graph.}
\label{CompletePlot}
\end{figure}

The basic concept here is to compare $p^{(f)}$ and $p^{(b)}$ to $r$, where $r = ( m/N)_{m=1}^{N}$. The sequence of $r_m$ represents the probabilities  corresponding to a coupon collector's problem. It is therefore known that
\begin{equation}\label{Comp-CollectToGumb}
S(r) \xrightarrow{d} G 
\end{equation}
where $G$ is distributed as $\mbox{ Gumb}(-\gamma, 1)$, as described in Section~\ref{Section_Star}.  

On the complete graph, we can rewrite $p_m$ (the probability of going from $m$ infected nodes to $m+1$ on the next time step) as $p_m = r_m ( 1 - \epsilon_m)$, where $\epsilon_m = (m-1)/(N-1)$. So for $N \gg m$, the $p_m$'s resemble the $r_m$'s quite closely. Therefore both the front tail of $p^{(f)}$ and the back tail of $p^{(b)}$ look suspiciously like coupon collector's processes. By this logic, we expect the total time to take over the complete graph should be just the sum of two coupon collector's times. That is, we suspect that $T$ is the sum of two Gumbel random variables. 

There are a few hangups with this intuitive argument:
\begin{enumerate}
\item $p^{(f)}$ and $p^{(b)}$ are about half the length of $r$.
\item $p_m$ doesn't quite equal $r_m$ at small $m$. 
\end{enumerate}
Addressing the first hangup involves, once again, the front tails of these vectors. Each $\mathcal{E}(p_m)$ has a standard deviation of $1/p_m$, which tells us that the smallest values of $p_m$ are the strongest drivers of the final distribution. 

The fact that the events at low populations (of either infected or susceptible types) strongly determine most of the random fluctuations is something that has shown up in other evolutionary models, especially with selective sweeps \cite{durrett04}. So if we were to just truncate both $p^{(f)}$ and $r$ at some point, we should expect the limiting distributions of $S(p^{(f)})$ or $S(r)$ to not substantially change. We formalize this idea in Appendix ~\ref{Truncation}, and find that it works out nicely.

We have a lot of options about where to truncate, but a useful truncation point is $B:= B(N) = \lfloor \sqrt{N-1} \rfloor$. The expression $\lfloor z \rfloor$ simply means we round $z$ down to the nearest integer. If we define
\begin{alignat*}{1}
p^{(T)} = (p_m)_{m=1}^B ,\hspace{0.5cm} \mbox{and } \hspace{0.5cm} r^{(T)} = (r_m)_{m=1}^B,
\end{alignat*}
then we have
\begin{alignat}{1}
S(p^{(f)}) \sim S(p^{(T)}), \hspace{0.5cm} S(p^{(b)}) \sim S(p^{(T)}) \label{Comp-HalfToTrunc}
\end{alignat} 
and 
\begin{alignat}{1}
S(r) \sim S(r^{(T)}). \label{Comp-CouponToTrunc}
\end{alignat} 

Addressing the second hangup mostly involves formalizing $\epsilon_m$ as a rather small number. The details are outlined in Appendix~\ref{Perturb}, where we find that 
\begin{equation} \label{Comp-TruncToTrunc}
S(r^{(T)}) \sim S(p^{(T)}).
\end{equation}
From here we can daisy-chain the previous numbered equations in this section together and find that 
\begin{equation}\label{CompleteDist}
\frac{T-\mu}{N} \xrightarrow{d} G+G.
\end{equation}
This means that we successfully piggy-backed on the result for star graphs to find that the resulting distribution for the complete graph is just a sum of two Gumbel random variables. The sum of two Gumbels has appeared previously in mathematically analogous places~\cite{aldous13, vanderhofstad2002flooding}. However our use of the coupon collector's problem makes for a quick conceptual justification.

Figure~\ref{CompletePlot} compares the takeover time distribution seen in simulations against the predicted distribution $\mathrm{Gumbel}(-\gamma, 1) \star \mathrm{Gumbel}(-\gamma, 1)$, and we see that this double-coupon logic works out well.

\section{d-dimensional lattice} \label{Section_dD}

As we did with the one-dimensional lattice, in our analysis of $d$-dimensional lattices we will assume periodic boundary conditions. The side length of the $d$-dimensional cube of $N$ nodes is denoted by $n = N^{1/d}$. We are also taking $1<d<\infty$, since we have already covered the 1D lattice and the infinite-dimensional lattice is a somewhat special case.

\begin{figure}
\includegraphics[width = 0.5\textwidth]{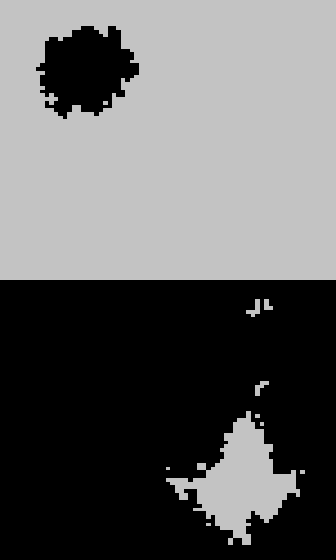}
\caption{Snapshots of our infection dynamics on a two-dimensional (2D) periodic cubic lattice. Black pixels show infected nodes, and grey pixels show susceptible nodes. The top panel shows a snapshot near the beginning of the dynamics, and the bottom panel shows a snapshot near the end. Notice how the blob of infected nodes in the top panel has a fairly simple shape, and most of the susceptible nodes lie in a single cluster in the bottom panel.}
\label{SpherePlot}
\end{figure}

Unlike every previous case we have examined, we cannot consistently define $p_m$. The probability of infecting a new node will almost always depend on the specific location of all currently infected nodes. This means that all our previous approaches will not work well here. However, this does not bar us from making guesses based on reasonable approximations.  

Although we could potentially get all kinds of weirdly-shaped clusters of infected nodes, that should not happen in expectation. Think back to the definition of our infection dynamics and Eq.~\eqref{pmLogic}. New infectees are added when a node on the boundary of the infected cluster gets randomly selected, and then one of its susceptible neighbors gets randomly selected and catches the infection. 

Intuitively, it sounds like we have an expanding blob of infected nodes, with the expansion happening uniformly outward on every unit of surface area. This is a recipe for making sphere-like blobs in $d$ dimensions, at least at the start of the dynamics. As seen from the top half of Figure~\ref{SpherePlot}, this looks plausible in two dimensions. 

The exact nature of this shape is actually a notoriously hard unsolved question. As we pointed out, there is a link between our infection model and first passage percolation on a lattice~\cite{auffinger15}. In that context, there is a rich literature surrounding questions about the nature of this cluster, but formal proofs of many of its properties have turned out to be difficult. However, convexity appears to be typical in the large size limit, and surface fluctuations should be relatively small~\cite{auffinger15}.  Moreover, there is good reason to believe that on the two-dimensional (2D) lattice, the boundary of the expanding cluster is a one-dimensional curve, which will come in handy later~\cite{bramson80}.  

In any case, since the lattice is periodic, this infected cluster will keep expanding. This means that at the end of the dynamics, we should expect for the majority of \emph{susceptible} nodes to also be in a single cluster, with insignificant enclaves elsewhere. This is borne out in simulations, as shown in the bottom half of Figure~\ref{SpherePlot}. If we focus on this majority susceptible cluster, we see that the end of the dynamics looks like a uniformly shrinking cluster of susceptible nodes, which is approximately the reverse of the uniformly growing infected cluster at the start. So, the beginning and end of the dynamics look similar once again, as they did for the complete graph.

More importantly, since this is a $d$-dimensional lattice, we can guess the surface area of these blobs. For a shape with a length-scale of $R$, we typically expect volume to scale as $R^d$ and surface area to go as $R^{d-1}$. So given an infected cluster of $m$ nodes, we expect it to have a surface area proportional to $m^{(d-1)/d}$. Assuming some uniformity, we should get that the typical probability of infecting a new node should be proportional to $m^{\eta}/N$ at the start of the dynamics, where the exponent $\eta$ is given by 
\begin{equation} \label{eta_equation}
\eta = \frac{d-1}{d}. 
\end{equation}
And just as in the case of the complete graph, this process at the start gets repeated backwards at the end. 

This heuristic argument suggests that the total time to takeover should look like the sum of geometric variables $X(p_m)$, where 
\begin{equation} \label{pm_d_dimensional}
p_m \approx \frac{m^{\eta}}{N} \left(1 - \frac{m^\eta}{N}\right).
\end{equation} 
The fact that we only got a grip on $p_m$ up to a proportionality should not worry us. After all, that did not stop us when we worked through the star graph case earlier; back then we argued that such a proportionality constant would simply show up in the scaling factor in the denominator. 
If we treat this as a numerical problem, we do not need to explicitly find the scaling factor. Instead, we can examine $(T - \mu)/\sigma$, where $\mu$ and $\sigma$ are empirically obtained values for the average and standard deviation of $T$ respectively. Then any proportionality constants just get absorbed by the anonymous $\sigma$.

\begin{figure}
\includegraphics[width = 0.5\textwidth]{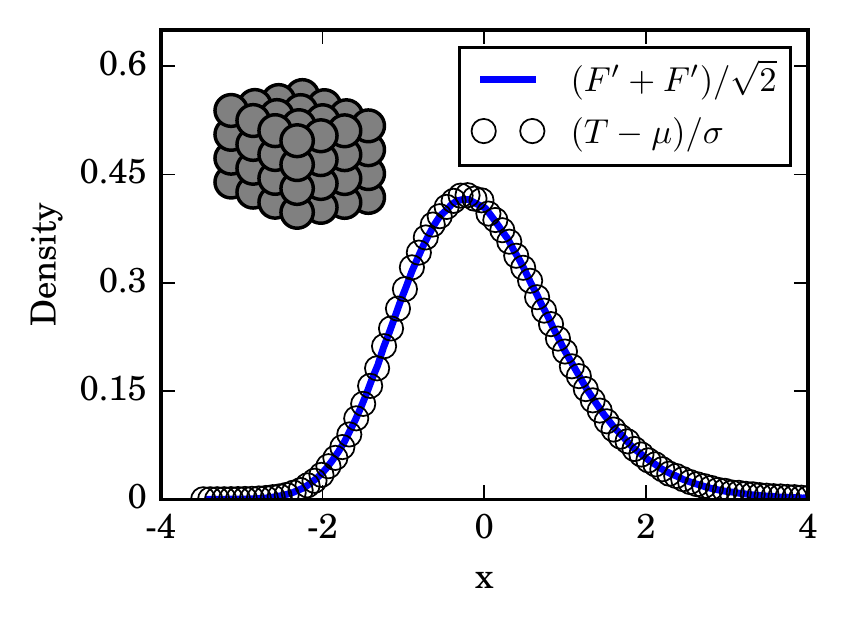} 
\caption{Distribution of takeover times $T$ for a 3D lattice with a side length of $n = 15$. The numerically generated distribution is based on $10^6$ simulation runs. The solid line shows the distribution of $(F'+F')/\sqrt{2}$, with $F'$ being summed up to $M=40$ and using $5 \times 10^6$ repetitions. The empirical quantities $\mu$ and $\sigma^2$ are the numerically calculated mean and variance of $T$. The schematic diagram in the upper left shows a 3D lattice.}
\label{3DLatticePlot}
\end{figure}

\begin{figure}
\includegraphics[width = 0.5\textwidth]{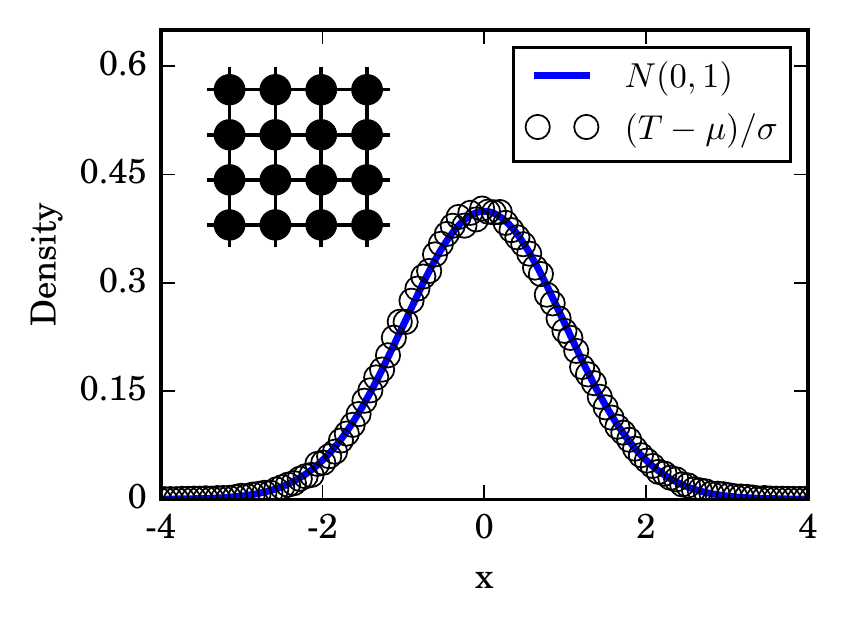} 
\caption{Distribution of takeover times $T$ for a 2D lattice with a side length of $n = 100$. The numerical results are obtained from $1.5 \times 10^5$ simulations. The solid line is the standard normal distribution. The empirical quantities $\mu$ and $\sigma^2$ are the numerically calculated mean and variance of $T$. The schematic diagram in the upper left shows a 2D lattice.}
\label{2DLatticePlot}
\end{figure}

This reasoning further suggests that for $N$ sufficiently large, 
\begin{equation}\label{LatticeToGeo}
\frac{T - \mu}{\sigma} \sim \sum_{m=1}^N\frac{ X( p_m ) - 1/p_m}{\sigma_X},
\end{equation}
where $\sigma_{X}^2 := \sum_{m=1}^N ( p_{m}^{-2} - p_{m}^{-1})$ is just the variance of the sum of geometric variables. But we already know how to approximate sums of geometric random variables. We can follow a similar procedure of truncation and perturbation as in the case of the complete graph. Assuming Eq.~\eqref{LatticeToGeo} is correct, we get 
\begin{equation} \label{LatticeToExp}
\frac{T - \mu}{\sigma} \sim \frac{1}{\sqrt{2}} (F' + F')
\end{equation}
where we define 
\begin{equation}\label{LatticeExpSum}
F' := \sum_{m=1}^M \frac{\mathcal{E}(m^\eta) - 1/m^{\eta}}{\sqrt{H}} 
\end{equation}
and 
\begin{equation} \label{H_equation}
H := H(2\eta; M) = \sum_{m=1}^M 1/m^{2\eta}
\end{equation}
for the sum of variances. 

The truncation point $M$ is some increasing function of $N$, which can normally just be set to $M=N$. In the limit of large $N$, the distinction does not really matter. However, it seems frequently possible to tune $M$ to get a good fit on finite-$N$ cases, as the simulations of the 3D lattices in Figure~\ref{3DLatticePlot} suggest. 

In principle, we could try to use Eq.~\eqref{ExpConv} to get a finite-$N$ estimate for this distribution. However, we do not expect any of these distributions to have a large-$N$ limit as easy as in the case of the star graph, nor for any of these distributions to have a name. For practical purposes, we can just simulate the right hand side of Eq.~\eqref{LatticeExpSum} directly, since generating and adding a large number of exponential variables is rather fast. 

\subsection*{The Critical Dimension} 

Naively, we might expect the limiting distribution of $F'$ to always be something between a Gumbel and a normal distribution. After all, $d = 1$ implies $\eta = 0$, which returns us to identical variables and the 1D ring, giving us the standard normal. Meanwhile, $d \to \infty$ implies $\eta \to 1$, which returns us to the coupon collector's problem and the star graph, giving us the Gumbel. Incidentally, this argument suggests that the infinite-dimensional lattice has similar behavior as the complete graph under these dynamics. In between these extreme cases, we might expect the intermediate $d$'s to correspond to a family of intermediate distributions. 

While this is generally true, there is a surprising caveat to be made about the case of $d=2$. Even though all the summands $(\mathcal{E}(m^\eta) - 1/m^{\eta})/\sqrt{H}$ are distinct, they start to resemble each other once $N$ gets sufficiently large. 

For $d = 2$, Eq.~\eqref{eta_equation} gives $\eta = 1/2$, which means that $H$ in Eq.~\eqref{H_equation} is the harmonic series. This $H$ diverges with $N$, giving each summand $(\mathcal{E}(m^\eta) - 1/m^{\eta})/\sqrt{H}$ a large denominator, and thus a small variance about a mean of zero. So, even though the summands are not identical random variables, they will become rather similar as we take $N$ to be large, suggesting that an improved version of the central limit theorem may apply. This intuition is confirmed by a careful analysis in Appendix~\ref{LatticeLindeberg}, showing that the Lindeberg-Feller theorem applies in this case. 

Thus we predict a normal limiting distribution of $F'$ in the specific case of the 2D lattice: as $N \rightarrow \infty$,
\begin{equation} \label{2DLatticeDist}
\frac{T - \mu}{\sigma} \xrightarrow{d} \mbox{Normal}(0,1) 
\end{equation}
for $d=2$. This prediction is borne out in simulation, as shown in Figure~\ref{2DLatticePlot}. 

However, no dimension higher than $d = 2$ can yield normally distributed takeover times. For each of $d = 3, 4, 5, \ldots$, the distribution of $F'$ will converge to a distinct limiting distribution between a normal and a Gumbel, as we initially suspected. The important distinction between $d = 2$ and $d > 2$ is that in the latter, $H$ always converges to a finite number. Because of that, $F'$ will always have a nonzero third moment, preventing it from converging to a standard normal. For more details, see Appendix~\ref{NonNormal}.

\section{Erd\H{o}s-R\'{e}nyi random graph} \label{Section_Erdos} 

Unlike all the previous graphs we have seen, an Erd\H{o}s-R\'{e}nyi graph is randomly constructed. We start off with $N$ nodes, and add an edge between any two with some probability $0< \rho \leq 1$. In this section, we will condition on the graph being connected, so that complete takeover is always possible. 

There is a good history of using generating functions to analyze desired properties on a random graph, including for various infection models~\cite{newman01, newman02, newman99}. But since we just finished analyzing the general lattice case, we can take another road. 

Recall the central observation that let us recast $T$ as a sum of geometric random variables. That train of logic only really involved the graph having a well-defined dimension $d$. If we could define the dimension for other kinds of graphs, then all our observations from the previous section would simply carry over. 

Imagine taking a cluster of $m$ nodes on an Erd\H{o}s-R\'{e}nyi graph. What is the surface area of said cluster? Well, in expectation, the $m$ nodes are externally connected to $O(pN)$ nodes, for $m \ll N$. So as $N$ gets large, the number of external neighbors in any cluster gets large as well, regardless of $m$. This is suggestive of an infinite-dimensional topology. 

So, by collecting results from Eqs.~\eqref{LatticeToExp} and~\eqref{CompleteDist}, we can guess the limiting distribution of the takeover times $T$. Defining $\mu$ and $\sigma$ to be the empirical mean and standard deviation of $T$, we find
\begin{equation}\label{ErdosDist}
\frac{T - \mu}{\sigma} \sim G' + G', 
\end{equation}
where $G'$ is a Gumbel random variable with a mean of zero and a variance of 1/2. One can check that the corresponding distribution for $G'$ is $\mathrm{Gumbel}(-\gamma\sqrt{3}/\pi, \sqrt{3}/\pi)$. 

\begin{figure}
\includegraphics[width = 0.5\textwidth]{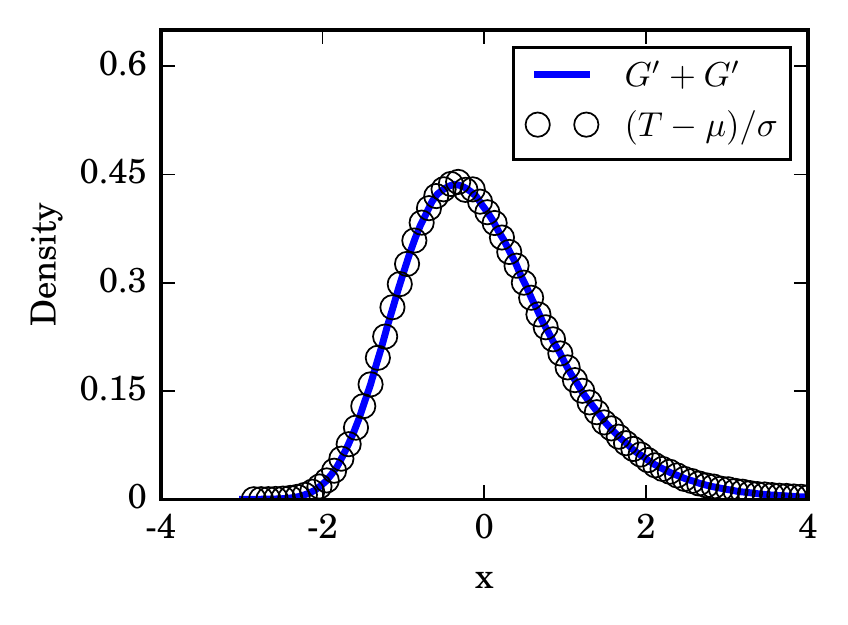} 
\caption{Distribution of takeover times $T$ for an Erd\H{o}s-R\'{e}nyi random graph on $N = 600$ nodes with an edge probability of $\rho = 0.5$. Simulation results were compiled from $10^6$ runs, all using the same realization of the random graph and all with the initial infection starting at the same node. The solid line was generated numerically by adding $5 \times 10^6$ pairs of Gumbel($-\gamma\sqrt{3}/\pi, \sqrt{3}/\pi$) random variables together. Similarly, $\mu$ and $\sigma^2$ are the numerically calculated mean and variance of $T$.}
\label{ErdosPlot}
\end{figure}

We experimentally tested Eq.~\eqref{ErdosDist} by fixing a randomly generated Erd\H{o}s-R\'{e}nyi random graph, along with a seed at which the infection always started. Then we ran a million simulations of the stochastic infection process and compiled the observed distribution of takeover times.  (The reason we fixed the graph beforehand was to avoid sampling multiple different values of $\mu$ and $\sigma$ over different realizations of the random graph.) The results of the experiment were consistent with our prediction, as shown in Figure~\ref{ErdosPlot}.

\section{Discussion} 

\subsection{Relation to other models}

\subsubsection{Infection models}

The model studied in this paper is intentionally simplified in several ways, compared to the most commonly studied models of infection. The purpose of the simplifications is to highlight how one aspect of the infection process -- its network topology -- affects the distribution of takeover times. However, the update rule also plays an important role. The assumptions we have made about it therefore deserve further comment.

Assumption 1: The infection is infinitely transmissible. When an infected node interacts with a susceptible node, the infection spreads with probability one. In a more realistic model, infection would be transmitted with a probability less than one. 

Assumption 2: The infection lasts forever. Once infected, a node never goes back to being susceptible, or converts to an immune state, or gets removed from the network by dying. The dynamics of these more complicated models, known as SIS or SIR, have been studied on lattices and networks by many authors; for reviews, see \cite{diekmann2012mathematical,pastor2015epidemic}. 

Assumption 3: The update rule is asynchronous. In other words, only one link is considered at a time. By contrast, in a model with synchronous updating, every link is considered simultaneously. 

If the infection is further assumed to be infinitely transmissible, then at each time step every infected node passes the infection to every one of its susceptible neighbors. Such a infection, akin to the spreading of a flood or a wildfire, would behave even more simply than the process studied here. In fact, it would be too simple. The calculation of the network takeover time would reduce to a breadth-first search and its value would be bounded above by the network's diameter. Note, however, that if the infection has a probability less than one of being transmitted to susceptible neighbors (such as in the original 1-type Richardson model~\cite{richardson1973random}), the system becomes nontrivial to analyze~\cite{auffinger15, richardson1973random}.

\subsubsection{Models of evolutionary dynamics}

More recently, the field of evolutionary dynamics~\cite{nowak2006evolutionary} has been extended to networks, and the field of evolutionary graph theory was born~\cite{lieberman05}. In general, the results in this field depend on modeling the spread of a mutant population using the Moran process~\cite{nowak2006evolutionary,moran58}. (Our model can be viewed as a limiting case of the Moran birth-death process, in the limit as the mutant fitness tends to infinity.) A number of important and interesting results have come from these studies of Moran dynamics, including the existence of network topologies that act as amplifiers of selection~\cite{adlam2015amplifiers}, increasing the probability of takeover, and also topologies that shift the takeover times we are considering~\cite{frean13}.

For example, working in the framework of evolutionary graph theory, Ashcroft, Traulsen, and Galla recently explored how network structure affects the distribution of ``fixation times'' for a population of $N$ individuals evolving by birth-death dynamics~\cite{ashcroft15}. The fixation time is defined as the time required for a fitter mutant (think of a precancerous cell in a tissue) to sweep through a population of less fit wild-type individuals (normal cells). Initially, a single mutant is introduced at a random node of the network. At each time step, one individual is randomly chosen to reproduce. With probability proportional to its fitness, it gives birth to one offspring, and one of its network neighbors is randomly chosen to die and be replaced by that offspring. The natural questions are: What is the probability that the lineage of the mutant will eventually take over the whole network? And if it does, how long does it take for this fixation to occur?  

The calculations are difficult because there is no guarantee of mutant fixation (in contrast to our model, where the network is certain to become completely infected eventually). In the birth-death model, sometimes by chance a normal individual will be chosen to give birth, and its offspring will replace a neighboring mutant. If this happens often enough, the mutant population can go extinct and wild-type fixation will occur. Using Markov chains, Hindersin and colleagues provided exact calculations of the fixation probability and average fixation times for a wide family of graphs, as well as an investigation of the dependence on microscopic dynamics~\cite{hindersin15, hindersin14, hindersin2016exact, hindersin2016should}. A challenge for this approach is that the size of the state space becomes intractable quickly: even with sparse matrix methods, it grows like $N2^{N}$~\cite{hindersin2016exact}. For networks of size $N < 23$, their computations showed that the distributions of mutant fixation times were skewed to the right, much like the Gumbels, sums of Gumbels, and intermediate distributions found analytically and discussed here in Sections \ref{Section_Star} to \ref{Section_Erdos}.

\begin{figure*}[t]
\includegraphics[width = 0.95\textwidth]{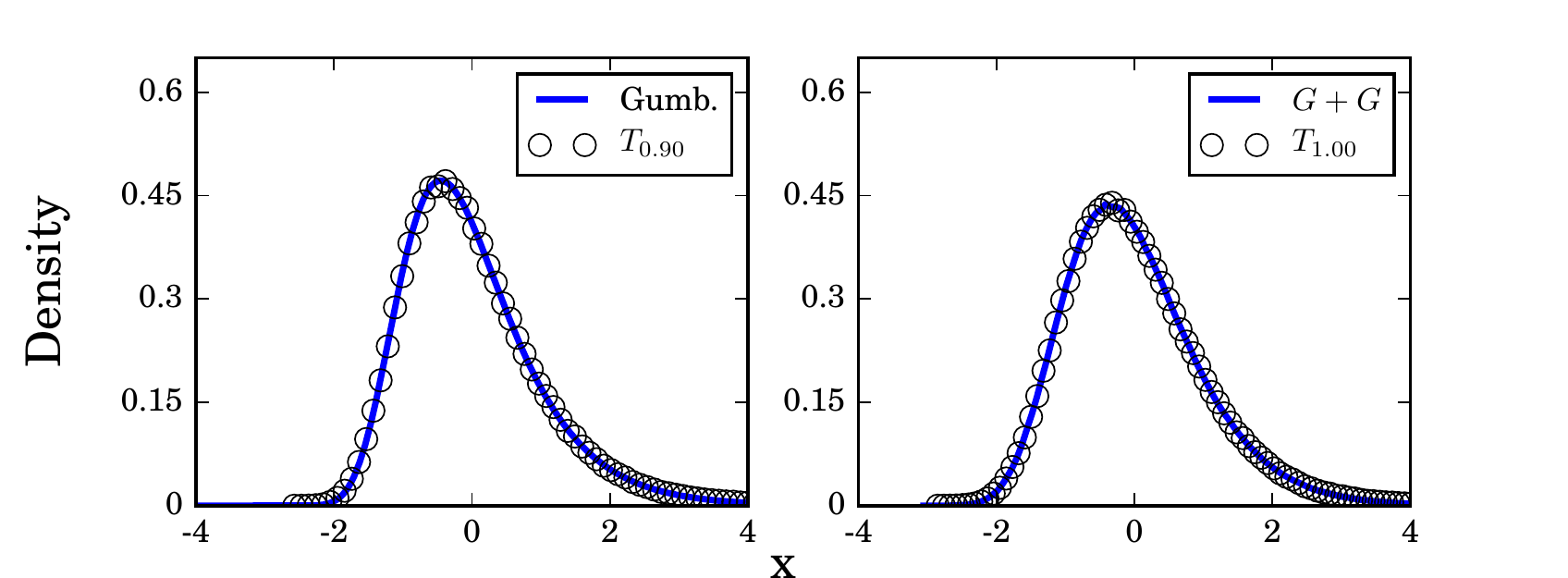} 
\caption{Normalized distribution of takeover times for an Erd\H{o}s-R\'{e}nyi random graph on $N = 600$ nodes with an edge probability of $\rho = 0.5$, obtained from $10^6$ simulation runs. For the sake of convenience, each $T$ is rescaled to have a mean of 0 and a variance of 1. The left panel shows the normalized times required to infect 90\% of the population, while the right shows the times for complete takeover. Here, the convolution of two Gumbel distributions plotted on the right was generated using $5 \times 10^6$ samples.}
\label{TruncPlot}
\end{figure*}

\subsubsection{First-passage percolation}

Our infection model is also closely related to first-passage percolation~\cite{aldous13,auffinger15}. The premise behind this family of models can be described as follows: given a network, assign a random weight to each edge. By interpreting that weight as the time for an infection to be transmitted across that edge, and by choosing properly tuned geometric (or, more commonly, exponential) random variables as the edge weights, we can recreate our infection model. 

Notice that percolation defines a random metric on the network, meaning that inter-node distances change from one realization to another.  This leads to a number of natural questions.  The most extensively studied is the ``typical distance,'' quantified by the total weight and number of edges on the shortest path between a pair of random nodes~\cite{bhamidi2013weak, eckhoff2015longI, eckhoff2015longII}. It is also possible to analyze the ``flooding time''~\cite{janson1999one,vanderhofstad2002flooding}, defined as the time to reach the \emph{last} node from a given source node chosen at random. This quantity is the closest analogue, within first-passage percolation, of our takeover time.  Indeed, a counterpart of our result for two Gumbels in the Erd\H{o}s-R\'enyi random graph was obtained previously using these techniques \cite{vanderhofstad2002flooding}.  However, we are unaware of flooding-time counterparts of our results about the takeover times for $d$-dimensional lattices.  

Another natural question in first-passage percolation involves finding the long-time and large-$N$ limiting shape of the infected cluster. More precisely, given a fixed origin node, we can identify all nodes that can be reached from the the origin within a total path-weight of $t$ or less. This amounts to finding all the nodes that have been infected by the origin within time $t$, a problem that percolation theorists have typically studied in $d$-dimensional lattices. We saw an instance of such an expanding cluster in Section \ref{Section_dD}.  In a general number of dimensions, the provable nature of this shape may be complicated; the fluctuations of its boundary are thought to depend on the KPZ equations~\cite{auffinger15, kardar1986dynamic,cieplak1996invasion}. The limiting shape is not typically a Euclidean ball, but it has been  proven to be convex; see \cite{aldous13,auffinger15} for an introductory discussion of these issues. In Figure~\ref{SpherePlot}, the nature of this cluster's complement in a large torus was of concern to us, but that issue has not yet attracted mathematical attention, as far as we know.

\subsection{Applications to medicine: epidemic and disease incubation times and cancer mortality} \label{medical}

For more than $100$ years, there have been intriguing empirical observations of ``right-skewed'' distributions in a remarkably wide range of phenomena related to disease~\cite{sawyer1914typhoid, miner1922incubation,sartwell1950distribution, sartwell1952incubation,sartwell66}. Examples include within-patient incubation periods for infectious diseases like typhoid fever~\cite{sawyer1914typhoid, miner1922incubation}, polio~\cite{sartwell1952incubation}, measles~\cite{goodall1931incubation} and acute respiratory viruses~\cite{lessler09}; exposure-based outbreaks like anthrax~\cite{brookmeyer2005modelling} (see~\cite{nishiura07,lessler09} for more recent reviews); rates of cancer incidence after exposure to carcinogens~\cite{armenian1974distribution}; and times from diagnosis to death for patients with various cancers~\cite{boag49} or leukemias~\cite{feinleib60}. 

The relationship between these phenomena and our model is intuitive: most of these processes depend on some sort of agent (a mutant cell, a virus, or a bacterium) invading and taking over a population, something which typically proceeds one  ``interaction'' at a time. And as we have seen, our simple infection model \emph{automatically} generates right-skewed distributions like Gumbels, sums of Gumbels, and intermediate distributions via a coupon-collection mechanism, for many kinds of population structures.  So could it be that the right-skewed distributions so often seen clinically are, at bottom, a reflection of this same mathematical mechanism, a manifestation of an invasive, pathogenic agent spreading through a network of cells or people?

To test the plausibility of this idea, we need to amend our model slightly. Until now we have focused exclusively  on the time $T$ to total takeover of a network. But in most scenarios related to disease, total takeover is not the relevant consideration. Sufficient takeover is what matters. For example, a patient need not have every single one of their bone marrow stem cells replaced by leukemic cells before they die from leukemia. Death presumably occurs as soon as some critical threshold is crossed -- which is probably the case for diseases with infectious etiologies as well. So let us now check whether changing the criterion from total takeover to partial takeover changes our results, or not.

\subsection*{Times to partial takeover: truncation} 

Define  $T_\theta$ to be the time for $\lfloor \theta  N \rfloor$ out of $N$ members to be infected, with the interesting range of $\theta$'s being $0.5 \leq \theta < 1.0$.  For the sake of example, consider the complete graph as our network topology, so we have $p_m = (m/N)(1 - (m-1)/(N-1))$.

As in the analysis for the complete takeover times, we can split $T_\theta$ into a front and back part $T^{(f)}$ and $T^{(b)}$, with the front covering up to about $N/2$ and the back covering the remaining. Then 
\begin{equation*}
\frac{T_\theta - \mu_\theta}{N} = \frac{T^{(f)} - \mu_{(f)}}{N} + \frac{\sigma_{(b)}}{N} \frac{T^{(b)} - \mu_{(b)}}{\sigma_{(b)}},
\end{equation*}

\noindent where $\mu_{(f)}$ is the mean of $T^{(f)}$, and $\mu_{(b)}$ and $\sigma_{(b)}$ are the mean and standard deviation of $T^{(b)}$.  However, it is easy to show that $\sigma_{(b)}^2 / N^2$ converges to 0 as $N$ gets large, regardless of $\theta$.  So we expect the distribution of $T_\theta$ in this case to asymptotically approach a Gumbel.  As seen in Figure~\ref{TruncPlot},  similar results hold for Erd\H{o}s-R\'{e}nyi random graphs, even for $\theta = 0.90$. 

Thus, for complete graphs and Erd\H{o}s-R\'{e}nyi random graphs, the right-skewed distributions for complete takeover persist when we relax the criterion to partial takeover. In that respect our results seem to be robust. 

The resilience of the Gumbel distribution is important to appreciate.  As pointed out by Read~\cite{read98}, a Gumbel distribution can be impersonated by a properly tuned 3-parameter lognormal distribution; see Appendix~\ref{lognormal_v_Gumbel} for further details. A 3-parameter lognormal distribution has a  density function
\begin{equation*}\label{lognormalDist}
h(x) = \frac{1}{(x-c)\sqrt{2\pi b^2}} \exp\left\{\frac{-\left[\log(x-c) - a \right]^2}{2b^2}  \right\}
\end{equation*}
provided $x > c$.  

It is this 3-parameter lognormal distribution that has been frequently noted in empirical studies of disease incubation times. Originally proposed and elaborated by Sartwell~\cite{sartwell1950distribution,sartwell1952incubation,sartwell66} as a curve-fitting model, its seeming generality has led to it being called ``Sartwell's Law.'' But it has always lacked a theoretical underpinning. Even recent reviews consider the origin of lognormal incubation times to be unresolved~\cite{nishiura07}. In contrast,  Gumbel and related distributions arise very naturally from the model studied here and from other infection models~\cite{williams65, gautreau07}, and may provide a more suitable theoretical foundation than lognormals in that sense. 

he spreading dynamics of processes on realistic, spatial topologies, Lieberman et al. introduced the field of `evolutionary graph theory'.

\appendix
\section{The Lindeberg condition (1D)} \label{RingLindeberg}

When we were analyzing the 1D lattice, we could not quite make use of the classical central limit theorem, since our variables $X$ have a dependence on $N$, being Geo($1/N$). Fortunately, there is the Lindenberg-Feller variant of the central limit theorem, which lets us get the desired normal convergence. See reference~\cite{durrett91}, page 98 for more details. However, we must first satisfy some special conditions before we can cite it. 

For convenience's sake, let us define 
\[ Y := \frac{X - N}{(N-1) \sqrt{N}}.\]
So $E[Y] = 0$, and $\sum_{m=1}^{N-1} E[Y^2] = 1$, which satisfies two of the three conditions. However, there is still the matter of the titular Lindeberg condition on the restricted second moments, which says for any fixed $\epsilon > 0$,

\begin{equation*}
\lim_{N\to\infty} \sum_{m=1}^{N-1} E[ Y^2 ; |Y| > \epsilon] = 0. 
\end{equation*}

To verify this, first notice that $|Y| > \epsilon$ means that we need either $X > N + \epsilon (N-1)\sqrt{N}$ or $X < N - \epsilon (N-1)\sqrt{N}$. However, the minus case will not come up in the limit; as $N$ gets large it would require $X$ to be negative, which is not possible. 

Letting $c: = \lfloor N + \sqrt{N}(N-1)\epsilon \rfloor$, we get
\begin{alignat*}{1}
&E[ Y^2 ; Y > \epsilon] \\
& = \sum_{k = c+1}^\infty \left(\frac{k-N}{(N-1)\sqrt{N}} \right)^2 \frac{1}{N} \left(1 - \frac{1}{N} \right)^{k-1} \\ 
& = \frac{1}{(N-1)N^2} \left( 1 - \frac{1}{N}\right)^c N \left[c^2 + N(N-1) \right]. 
\end{alignat*}
And so, 
\begin{alignat*}{1}
& \lim_{N \to \infty} \sum_{m=1}^{N-1} E[ Y^2 ; |Y| > \epsilon] \\
& = \lim_{N \to \infty} \frac{1}{N(N-1)} \left(1 - \frac{1}{N} \right)^c \left[c^2 + N(N-1) \right] \\
&= \lim_{N \to \infty} \left(1 - \frac{1}{N}\right)^c + \lim_{N \to \infty} \frac{c^2}{N(N-1)} \left(1 - \frac{1}{N} \right)^c . 
\end{alignat*}
Here, the first limit looks like 
\begin{alignat*}{1}
\lim_{N \to \infty} \left(1 - \frac{1}{N}\right)^{N - \epsilon N^{1/2} + \epsilon N^{3/2}} = 0. 
\end{alignat*}
Meanwhile the second limit can be bounded above (with some constant $C$) by 
\begin{alignat*}{1}
\lim_{N \to \infty} C N\left(1 - \frac{1}{N} \right)^{\epsilon N^{3/2}} = 0. 
\end{alignat*}
So the total sum of conditional expectations converges to 0 as $N$ gets large, and so the Lindeberg condition is satisfied. This allows us to cite the theorem, and confirms that in the limit we get 
\begin{equation}
\sum_{m=1}^{N-1} \frac{X - N}{(N-1) \sqrt{N}} \xrightarrow{d} \mbox{Normal}(0,1).
\end{equation}

\section{Geometric variables converging to exponential variables} \label{ConvergeToExp}

\noindent {\bf Proposition: } {\it
Say we have a positive sequence $(p_m)_{m=1}^M$, and some function $L := L(M)$ such that 
$\lim_{M\to \infty} L = \infty$
and 
\[\lim_{M \to \infty} \sum_{m=1}^M \frac{1}{p_mL^2} = 0.\] 
Then if $T := \sum_{m=1}^M X(p_m)$, $F := \sum_{m=1}^M \mathcal{E}(p_m)$, and $\mu := \sum_{m=1}^M 1/p_m$, we have
\begin{equation} \label{GeoToExp}
\frac{T - \mu}{L} \sim \frac{F - \mu}{L} .
\end{equation}
}

\noindent {\bf Proof:} This is proven by finding the characteristic functions for both sides, and showing that the ratio of these functions goes to 1 as $M$ gets large. The characteristic function of a random variable uniquely determines its distribution, so this is a rather powerful statement. 

Let us define
\begin{alignat*}{1}
\Phi := E\left[\exp\left( it \frac{T - \mu}{L} \right) \right] .
\end{alignat*}
If we split $T$ into the sum of geometric random variables and rearrange, we eventually get
\begin{equation} \label{GeoChar}
\Phi = \prod_{m=1}^{M} \frac{p_m \exp\left[( it/L) \left(1 - 1/p_m \right) \right] }{1 - q_m \exp\left( it/L \right)}.
\end{equation}
Similarly, if we set 
\begin{alignat*}{1}
\phi := E\left[\exp\left( it \frac{F -\mu}{L} \right) \right],
\end{alignat*}
then after we proceed through some more algebra, we find that 
\begin{equation} \label{ExpChar}
\phi = \prod_{m=1}^M \frac{\exp\left[ -it / (p_mL) \right]}{1 - it/(p_mL) }.
\end{equation}

Let us fix $t$ so that we can pointwise consider the ratio of the characteristic functions. After some manipulation, we find
\begin{equation*}
\phi/\Phi = \prod_{m=1}^M \frac{\exp(-it/L) -q_m}{p_m \left[1 - it/(p_mL) \right]}. 
\end{equation*}
We assumed that $L$ gets large, so there is some function $R_1 := R_1(M)$ that has vanishing magnitude with large $M$ such that 
\begin{equation*}
\exp(-it/L) = 1 +(-it/L) + R_1 t^2/L^2.
\end{equation*}
So then we have 
\begin{equation*}
\phi/\Phi = \prod_{m=1}^M \left( 1 + \frac{t^2}{p_mL^2} \frac{ R_1} { 1 - it/(p_mL) } \right). 
\end{equation*}
Notice that $|1 - it/(p_mL)| \geq 1$. In addition, we already know the sum of $1/(p_mL^2)$ goes to 0, so it must be that each individual $p_mL^2$ gets large for all $m$. This ensures the second term is small, and therefore it can be rewritten exactly as an appropriate exponential. 

So 
\begin{alignat*}{1}
\phi/\Phi =& \prod_{m=1}^M \exp\left[R_2 t^2 /(p_m L) \right] \\
=& \exp\left[ t^2 R_2 \sum_{m=1}^M \frac{1}{p_m L^2} \right] \to 1,
\end{alignat*}
where the final limit comes from our assumption on $\sum_{m=1}^M \frac{1}{p_m L^2}$. The limit converges to 1, which establishes the proposition. 

\section{Sum of exponentials} \label{ExpSum}
\noindent {\bf Proposition: } {\it If we have exponential random variables $\mathcal{E}(p_m)$ for $m = 1, \ldots , n$, with $p_m$ distinct, then $\sum_{m=1}^n \mathcal{E}(p_m)$ is distributed according to the density 
\begin{equation} \label{ExpConv}
g_n(x) = \sum_{k=1}^n p_k e^{-p_k x} \prod_{r = 1, r\not=k}^n \frac{p_r}{p_r-p_k}
\end{equation}
on $x \geq 0$. }

\noindent {\bf Proof:} 
This is a straightforward induction for the most part. The base case is simply checked by plugging in $n=1$. To get the inductive step down, we just convolve the previous step with a new exponential distribution, so 
\begin{equation*}
g_{n+1}(x) = \int_{0}^{x} p_{n+1} e^{-p_{n+1} (x-y)} g_n(y) dy .
\end{equation*}
After calculating for a bit, we find
\begin{alignat*}{1}
g_{n+1}(x) =& \sum_{k = 1}^n p_k e^{-p_k x}\prod_{ r\not=k}^{n+1} \frac{p_r}{p_r-p_k} \\
& + \sum_{k=1}^n \frac{p_k p_{n+1}}{p_k - p_{n+1}} e^{-p_{n+1} x} \prod_{r\not=k}^n \frac{p_r}{p_r-p_k}. 
\end{alignat*}
The first term is in the desired form, but the second term requires some work. After some further manipulation, we can get
\begin{alignat*}{1}
\mbox{Second Term} =& e^{-p_{n+1}x} \left( \prod_{k=1}^{n} \frac{p_k}{p_k - p_{n+1}} \right) b(p_{n+1}),
\end{alignat*}
where we define
\begin{equation*}
b(z) := \sum_{k=1}^n \prod_{r \not= k}^n \frac{p_r - z}{p_r - p_k}.
\end{equation*}
We can interpret $b(z)$ as a polynomial of at most degree $n-1$ in $z$ (a Lagrange polynomial, to be specific). 

But notice that for $l = 1, \ldots, n$, then $b(p_l) = 1$. This means that $b(z) - 1$ is a polynomial with n distinct roots, which is more than what its maximum degree should normally allow. The only way that is possible is if $b(z) - 1$ is a constant 0, so $b(z) \equiv 1$. Plugging this in and simplifying gives
\begin{equation*} 
g_{n+1}(x) = \sum_{k=1}^{n+1} p_k e^{-p_k x} \prod_{r = 1, r\not=k}^{n+1} \frac{p_r}{p_r-p_k},
\end{equation*}
which is the desired result.

\section{Truncation of sequences} \label{Truncation}

\noindent {\bf Proposition: } {\it Let $L := L(M)$, $B := B(M)$ with $\lim_{M\to\infty} L(M) = \lim_{M \to \infty} B(M) = \infty$. Further say that $B$ is integer valued with $1\leq B \leq M$. Given a positive sequence $p(M) = (p_m)_{m=1}^M$, assume
\begin{equation*}
\lim_{M \to \infty} \sum_{m=1}^M \frac{1}{(p_mL)^2} = A < \infty,
\end{equation*}
and $\lim_{M \to \infty} 1/[p_{k(M)}L] = 0$ given that $M \geq k(M) > B$. Then
\begin{equation} \label{TruncEq}
\sum_{m=1}^B \frac{\mathcal{E}(p_m) - 1/p_m}{L} \sim \sum_{m=1}^M \frac{ \mathcal{E}(p_m) - 1/p_m }{L}.
\end{equation}}
\noindent {\bf Proof:} 
As before, the proof involves showing the ratio of characteristic functions converges to 1. The full series on the right has the function 
\begin{equation*}
\phi = \prod_{m=1}^M \frac{\exp[-it/(p_mL)]}{ 1 - it/(p_mL)},
\end{equation*}
and that the truncated series on the left has
\begin{equation*}
\hat\phi = \prod_{m=1}^B \frac{\exp[-it/(p_mL)]}{ 1 - it/(p_mL)}.
\end{equation*}
So naturally, we fix a $t$ and get the ratio
\begin{equation*}
\phi/\hat\phi = \prod_{m=B+1}^M \frac{\exp[-it/(p_mL)]}{ 1 - it/(p_mL)}.
\end{equation*}

Because of the our last condition, we know that $1/(p_mL)$ is small for all $m$ in this range. So we can do a Taylor expansion and make a function $R_1$ which is small in magnitude so that 
\begin{equation*}
\phi/\hat\phi = \prod_{m=B+1}^M \left(1 + \frac{t^2}{(p_mL)^2} \frac{ R_1 }{1 - it/(p_mL)} \right) .
\end{equation*}
Again, $| 1 - it/(p_mL) | > 1$ and $p_mL$ is large, so we can again shift to an exponential to get 
\begin{equation*}
\phi/\hat\phi = \exp\left[ \sum_{m=B+1}^M \frac{R_2 t^2} {(p_mL)^{2}} \right],
\end{equation*}
where $R_2$ is small in magnitude again. But notice that this is based on the tail of a convergent sum. So
\begin{alignat*}{1}
&\lim_{M\to\infty} \sum_{m=B+1}^M \frac{1}{(p_mL)^{2}} \\
&= \lim_{M\to\infty}\left( \sum_{m=1}^M \frac{1}{(p_mL)^{2}} - \sum_{m=1}^B \frac{1}{(p_mL)^{2}} \right) \\
& = A - A = 0.
\end{alignat*}
And so $\lim_{M \to \infty} \phi/\hat\phi = 1$.

\section{Edge perturbations} \label{Perturb}

In principle we could show a more general statement here, but we are only going to directly calculate the effect of a perturbation once in this paper. So, for the sake of readability, we are just going to do this specific example. 

Recall that for the complete graph $Np_m = m ( 1 - (m-1)/(N-1) ) = N r_m ( 1 - \epsilon_m)$ with $r_m = m/N$. Also recall that $p^{(T)}$ and $r^{(T)}$ are the truncated sequences up to $B = \lfloor \sqrt{N-1} \rfloor$. Let $\hat \phi$ be the characteristic function associated with the normalized sum $S(p^{(T)})$, and that $\phi$ is the characteristic function associated with $S^{(T)}$. Then
\begin{alignat*}{1}
\phi/\hat\phi =& \prod_{m=1}^B \exp\left[ \frac{-it}{r_mN} + \frac{it}{p_m N} \right] \frac{1 - it/(p_m N)}{1 - it/( r_m N)} \\ 
=& \prod_{m=1}^B \exp\left[ \frac{-it}{m} \left( 1 - \frac{1}{1-\epsilon_m} \right) \right] \\
&\times \frac{1 - it(1 - \epsilon_m)^{-1} /m}{1 - it/m }. 
\end{alignat*}
Notice $\epsilon_m = (m-1)/(N-1) = O\left(N^{-1/2}\right)$ in the range of $m$'s in the product, and is therefore small. We go through some Taylor expansions and cancellations, and using $R_j$ to represent functions of small magnitude, we find
\begin{equation*}
\phi/\hat\phi = \prod_{m=1}^B\left( 1 + \frac{ it R_2 \epsilon_m }{ m } \right) \exp\left( \frac{itR_1\epsilon_m}{ m } \right).
\end{equation*}
The first term can be once again turned into an exponential (thanks to the smallness of $\epsilon$), and so we get
\begin{equation*}
\phi/\hat\phi = \exp\left(2 it R_3 \sum_{m=1}^B \frac{\epsilon_m}{ m } \right).
\end{equation*}
Therefore, we get convergence to 1 if the sum converges to 0 as $N$ gets large. But this sum is easy to bound from above. That is, 
\begin{alignat*}{1}
\sum_{m=1}^B \frac{\epsilon_m}{ m } =& \sum_{m=1}^{\lfloor \sqrt{N-1} \rfloor} \frac{1}{ m } \frac{m-1}{N-1} \\
\leq & \sum_{m=1}^{\lfloor \sqrt{N-1} \rfloor} \frac{1}{ m } \frac{m}{N-1} \\
& \leq (N-1)^{-1/2} \xrightarrow{N \to \infty} 0.
\end{alignat*}

This means we get that $\phi/\hat\phi \to 1$, implying that the truncated sum for the complete graph converges to the truncated distribution for the coupon collector's problem.

\section{The Lindeberg condition (2D)} \label{LatticeLindeberg}

Much like in the 1D lattice case, we are unable to directly use the typical central limit theorem, because the variables are not identical and have a dependence on $N$. But once again, we can apply the Lindenberg-Feller theorem. We are going to focus on the 2D case, so we have $\eta = 1 - 1/d = 1/2$. Let
\begin{equation*}
Y_{N,m} = \frac{\mathcal{E}(\sqrt{m}) - 1/\sqrt{m}}{\sqrt{H}}. 
\end{equation*}
Because we are only looking at the special 2D case, $H = H(N, 2(1/2)) = \sum_{k=1}^N1/k$. 

Notice for any higher dimension $d > 2$, then we would get $H = \sum_{k=1}^N 1/k^{2-2/d}$, which quickly converges to a finite number as $N$ gets large, whereas with $d =2$, we have $H$ gets large for large $N$. This distinction is what lets us apply the theorem to the 2D case, but not the rest. 

Anyway, it is easy to check that $E[Y_{N,m}] = 0$ and $E[Y_{N,m}^2] =m^{-2\eta}/ H$. So then 
\begin{equation*}
\sum_{m=1}^N E[Y_{N,m}^2] = \frac{1}{H} \sum_{m=1}^N \frac{1}{m} = 1.
\end{equation*}
So in order to apply the theorem, we only need to check if for any fixed $\epsilon > 0$, we have 
\begin{equation} \label{2DLindCond}
\mbox{Lind.}:= \lim_{N \to \infty} \sum_{m=1}^N E[Y_{N,m}^2; |Y_{N,m}| > \epsilon] \stackrel{?}{=} 0.
\end{equation}
If this final condition holds, then we can cite the theorem and conclude that $\sum_{m=1}^N Y_{N,m}$ is distributed as a normal as $N$ gets large.

We need not care about the $ Y_{N,m} < -\epsilon$ case, because this is equivalent to asking for $\mathcal{E}(\sqrt{m}) < 1/\sqrt{m} - \epsilon \sqrt{H}$. However, $H$ scales as $\log(N)$ in the limit of large $N$ whereas $1/\sqrt{m} \leq 1$, so this quantity will always eventually become negative, whereas exponential variables are always positive. 

Therefore, let us focus on the positive half. Letting $c := 1/\sqrt{m} + \epsilon \sqrt{H} $ and integrating, we get
\begin{alignat*}{1}
& E[Y_{N,m}^2; Y_{N,m} > \epsilon] \\
&= \int_{c}^{\infty} \sqrt{m} e^{-\sqrt{m} x} \left(\frac{x - 1/\sqrt{m}}{\sqrt{H} }\right)^2 dx \\
&= \frac{1}{H m } \left( 1 + (\sqrt{m} c)^2 \right) e^{-\sqrt{m} c}.
\end{alignat*}
Substituting in gives us
\begin{alignat*}{1}
& E[Y_{N,m}^2; Y_{N,m} > \epsilon] \\
&= e^{-1} \left( \frac{2}{H m} + \frac{2\epsilon} {\sqrt{m} \sqrt{H}} + \epsilon^2 \right) \exp(-\epsilon \sqrt{m} \sqrt{H} ).
\end{alignat*}
That last term will be the dominant term as $N$ gets large, so we can choose some positive constant $C_1$ (which may depend on $\epsilon$) such that
\begin{alignat*}{1}
& E[Y_{N,m}^2; Y_{N,m} > \epsilon] \leq C_1 \exp(-\epsilon \sqrt{m} \sqrt{H} ).
\end{alignat*}
So, we have 
\begin{alignat*}{1}
\mbox{Lind.} 
&= \lim_{N \to \infty} \sum_{m=1}^N E[Y_{N,m}^2; Y_{N,m} > \epsilon] \\
& \leq \lim_{N \to \infty} \sum_{m=1}^N C_1 \exp(-\epsilon \sqrt{m} \sqrt{H} ).
\end{alignat*}
We can bound the harmonic sum $H$ from below with a constant times $\log(N)$, so there is some positive $C_2$ such that 
\begin{equation*}
\mbox{Lind.} \leq \lim_{N \to \infty} \sum_{m=1}^N C_1 \exp\left(-C_2 \sqrt{\log(N)} \sqrt{m}\right).
\end{equation*}
We can approximate this sum from above by interpreting it as a Riemann sum. By taking the appropriate integral, we get 
\begin{alignat*}{1}
& \int_{0}^N C_1 \exp\left(-C_2 \sqrt{\log(N)} \sqrt{x}\right)dx \\
=& \frac{2C_1}{C_2\log(N)} \left[1 - \exp\left(-C_2 \sqrt{N\log(N)} \right) \right. \\
& \left. - C_2 \sqrt{N\log(N)} \exp\left(-C_2 \sqrt{N\log(N)} \right) \right] \\
\leq & \frac{2C_1}{C_2\log(N)}.
\end{alignat*}
Second moments are always nonnegative, which means
\begin{equation*}
0 \leq \mbox{Lind.} \leq \lim_{N \to \infty} \frac{2C_1}{C_2\log(N)} = 0.
\end{equation*}
So Eq.~\eqref{2DLindCond} is finally confirmed. As a consequence, we can finally cite the Lindeberg-Feller theorem, and know that 
\begin{equation}\label{2DDistAppendix}
\sum_{m=1}^N \frac{\mathcal{E}(\sqrt{m}) - 1/\sqrt{m}}{\sqrt{H}} \xrightarrow{d} \mbox{Normal}(0,1).
\end{equation}

\section{Non-Normality of $d > 2$} \label{NonNormal}

There are a lot of possible ways to show a distribution {\it does not} converge to a normal in a limit. But to show that the distribution of $F'$ for a $d>2$ dimensional lattice (as defined in Eq.~\eqref{LatticeExpSum}) is not normal, it will suffice to consider the moments. We already know that $F'$ has a mean of 0 and a variance of 1; so if $F'$ went like a normal, then we should expect that $\lim_{N\to\infty}E[F'^3] = 0$ by symmetry. 

We can reuse Eq.~\eqref{ExpChar} to find the characteristic function $\phi$ of $F'$ by plugging in $p_m = m^\eta$ and $L^2 = H(2\eta) = \sum_{m=1}^N m^{-2\eta}$. Because $d > 2$, then $2\eta > 1$.   By the definition of the characteristic function, we know that if we expand $\phi$ in powers of $t$, then
\begin{alignat*}{1}
\phi =& 1 - it E[F'] - \frac{t^2}{2} E[{F'}^2 ] + i \frac{t^3}{6} E[{F'}^3] + \mbox{h.o.t.} \\
=& 1 - \frac{t^2}{2} + i \frac{t^3}{6} E[{F'}^3] + \mbox{higher order terms}.
\end{alignat*}
So we can get the third moment by just reading off the coefficient of the $t^3$ term. 

Returning to equations~\eqref{LatticeExpSum} and~\eqref{ExpChar}, let $x_m = it m^{-\eta}H(2\eta)^{-1/2}$. Using the standard expansions for $e^x$ and $1/(1-x)$, we find
\begin{alignat*}{1}
\phi =& \prod_{m=1}^N \frac{\exp(-x_m)} {1 - x_m} \\ 
=& \prod_{m=1}^N \frac{1 - x_m + \sum_{k=2}^\infty (-x_m)^k/k! }{1 -x_m} \\ 
=& \prod_{m=1}^N\left[1 + \left(\sum_{l=0}^{\infty} x_{m}^l \right) \left( \sum_{k=2}^\infty (-x_m)^k/k! \right) \right] \\
=& \prod_{m=1}^\infty \left[1 + \frac{x_{m}^2}{2} + \frac{x_{m}^3}{3} + \mbox{h.o.t.} \right].
\end{alignat*}
If we do not care about high order terms in $t$, then this is an easy product to take. In fact, if we collect terms and plug in for $x_m$, we get
\begin{alignat*}{1}
\phi =& 1 - \frac{t^2}{2H(2\eta)}\sum_{m=1}^{N} \frac{1}{m^{2\eta}} \\
&+ \frac{-i t^3}{3H(2\eta)^{3/2}} \sum_{m=1}^{N} \frac{1}{m^{3\eta}} + \mbox{h.o.t.} \\
=& 1 - \frac{t^2}{2} -i \frac{t^3}{3} \frac{H(3\eta)}{H(2\eta)^{3/2}} + \mbox{h.o.t.}.
\end{alignat*}
This means for any finite $N$, the third moment of $F'$ is simply
\begin{equation*}
E[{F'}^3] = -2\frac{H(3\eta)}{H(2\eta)^{3/2}}.
\end{equation*}

Although $H$ is a function that depends on $N$, this quantity will never get large. In fact, since $1 \geq \eta > 1/2$, then we know in the limit of large N that 
\begin{equation} \label{ThirdMoments}
E[{F'}^3] \to -2\frac{\zeta(3\eta)}{\zeta(2\eta)^{3/2}},
\end{equation}
where $\zeta$ is the Riemann zeta function. In the range of $\eta$'s presented, $\zeta$ neither diverges nor hits zero, so the above will never be zero. In fact, the right hand side of Eq.~\eqref{ThirdMoments} is monotone, so each distinct $1\geq \eta > 1/2$ will produce a distinct third moment, and therefore a distinct distribution.  As a side note, if we take $\eta \to 1$, we get $-12\sqrt{6}\zeta(3)/\pi^3$, which is the correct value for the third moment of a normalized Gumbel distribution, as expected. However, since the rest are distinct, that means we only transition to an exact Gumbel in the extreme limit.  

In summary: given $d>2$, we never expect $F'$ to have a zero third moment in the limit of large $N$, and so $F'$ can never converge to a normal distribution. Moreover, because their third moments depend on $\eta = 1 - 1/d$, we expect $F'$ to converge to a different distribution for each $d > 2$. Hence we expect there is no upper critical dimension.

\begin{figure*}
\includegraphics[width = 0.95\textwidth]{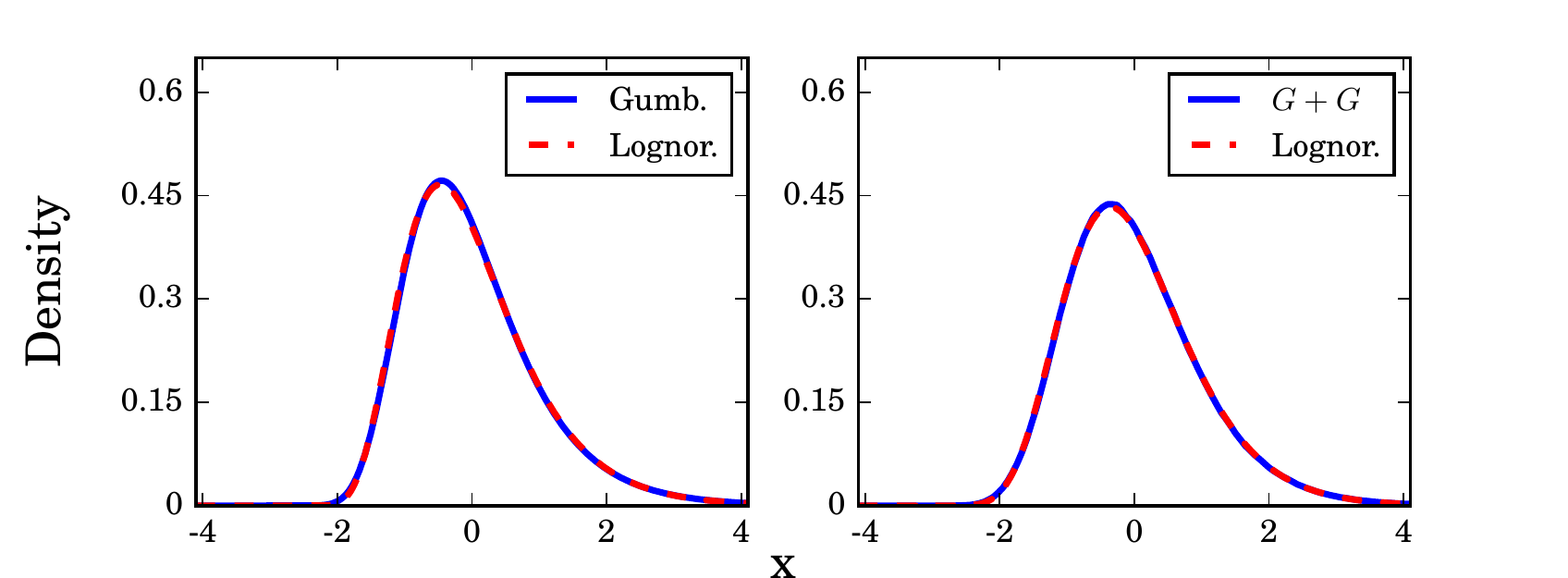} 
\caption{A properly chosen 3-parameter lognormal distribution can closely approximate a Gumbel or a sum of two Gumbel distributions. The parameters in these lognormals were chosen to fit the first three moments of the Gumbel or Gumbel+Gumbel distributions.}
\label{Gumbel_v_lognormal}
\end{figure*}
\section{Lognormal distributions can masquerade as Gumbels}
\label{lognormal_v_Gumbel}

In Section \ref{medical} we noted that the distribution of disease incubation periods and other times of medical interest  have often been fit by a lognormal. However, given the noise in real data, it is entirely possible that the true distribution should have been be a Gumbel (or a sum of two Gumbels), and was impersonated by a similar-looking lognormal.

Moreover, since most studies used 3-parameter lognormals, it would always be possible to match the first three moments of the data.  We do as such in Figure~\ref{Gumbel_v_lognormal}, producing a very close fit to both a Gumbel and a convolution of two Gumbels.  We can compare these densities using the Kolmogorov metric, given by the maximum difference between their cumulative distribution functions. Using this, we find that the normalized Gumbel is $\approx0.0034$ away from its corresponding lognormal in this metric.  For the sum of two Gumbels, we can numerically estimate that its corresponding lognormal is $ \lesssim 10^{-2}$ away in the Kolmogorov metric.


\section*{Acknowledgments}

Thanks to David Aldous, Rick Durrett, Remco van der Hofstad, Lionel Levine, and Piet Van Mieghem for helpful conversations. This research was supported by a Sloan Fellowship and NSF Graduate Research Fellowship grant DGE-1650441 to Bertrand Ottino-L\"{o}ffler in the Center for Applied Mathematics at Cornell, and by NSF grants DMS-1513179 and CCF-1522054 to Steven Strogatz. Jacob Scott acknowledges the NIH for their generous loan repayment grant. 

\bibliography{SITakeover_Bib}{}
\bibliographystyle{apsrev4-1-prx}

\end{document}